\documentclass[1p,times,preprint]{elsarticle}

\usepackage[english]{babel}
\usepackage{tensor}
\usepackage{graphicx}
\usepackage{amsmath}
\usepackage{amssymb}
\usepackage{amsfonts}
\usepackage{dcolumn}
\usepackage{bm}
\usepackage{xcolor}
\usepackage{ulem}
\usepackage{tikz}
\usepackage{subcaption}
\usepackage{comment}
\usepackage{verbatim}
\usepackage{fancyvrb}
\usepackage{cancel}
\usepackage{multirow}
\usepackage{lscape}
\usepackage{txfonts}
\usepackage{mathtools}
\usepackage{soul}
\usepackage{url}
\usepackage{makecell}
\usepackage[pdftex]{pict2e}
\usepackage{microtype}

\def\bra<#1|{\mathinner{\langle\,{#1}\,\vert}} 
\def\ket|#1>{\mathinner{\vert\,{#1}\,\rangle}} 
\def\red|#1|{\mathinner{\!\vert\,{#1}\,\vert\!}}
\def\braket<#1>{\mathinner{\langle\,{#1}\,\rangle}} 

\def\redmem#1#2#3{  \left\langle #1 \left\Vert  
                  #2 \right\Vert #3 \right\rangle   }

\begin{document}

\title{SECOND-ORDER RAYLEIGH-SCHR\"ODINGER PERTURBATION THEORY FOR THE GRASP2018 PACKAGE: CORE CORRELATIONS}

\author[TFAI]{G. Gaigalas}
\address[TFAI]{Institute of Theoretical Physics and Astronomy, 
               Vilnius University, Saul\.{e}tekio Ave. 3, LT-10257 Vilnius, Lithuania}
\ead{gediminas.gaigalas@tfai.vu.lt}

\author[TFAI]{P. Rynkun}
\ead{pavel.rynkun@tfai.vu.lt}

\author[TFAI]{L. Kitovien\.{e}}
\ead{laima.radziute@tfai.vu.lt}

%
%
\begin{abstract}
The paper presents a further development of the method G. Gaigalas, P. Rynkun, L. Kitovienė, 
Second-Order of Rayleigh-Schr\"odinger Perturbation Theory for the {\sc Grasp}2018 Package: Core-Valence
Correlations, {\em Lithuanian Journal of Physics}, 64, No. 1, 20-39 (2024) 

\hspace{-0.6cm} (https://doi.org/10.3952/physics.2024.64.1.3),
based on a combination of the relativistic configuration interaction method 
 and on the stationary second-order Rayleigh-Schr\"odinger many-body perturbation theory in an irreducible tensorial form.
In this extension the perturbation theory accounts for both electron
core-valence and core correlations when an atom or ion has any number of valence electrons
meanwhile relativistic configuration interaction accounts for the rest of correlations.
This allows a significant reduction of the space of the configuration state function
for complex atoms and ions. We also demonstrate how this method works
for energy structure calculation of Fe XV ion.

\end{abstract}


\maketitle


\section{Introduction}

A major challenge in atomic structure calculations is the accurate description of electron correlations.
Many powerful theoretical methods, such as different versions of many-body perturbation theory (MBPT) \cite{LindgrenBook:82,PT_book},
the configuration interaction method (CI) \cite{fischer_brage}, the relativistic configuration interaction (RCI) \cite{Fisetal:16a}, the random phase approximation
with exchange (RPAE) \cite{RPAE_1,RPAE_2}, the  multiconfiguration Hartree-Fock method (MCHF)
\cite{fischerBook:77} or the multiconfiguration Dirac-Hartree-Fock (MCDHF) method \cite{grantBook:07} 
have been developed to account for correlation effects.
These methods have their own disadvantages in obtaining highly accurate atomic data.
For example, the inclusion of correlation effects in the MCHF, MCDHF, or CI methods 
rapidly increase the expansion of atomic state function (ASF), especially for complex atoms.
Perturbation theory (PT) has practical and theoretical difficulties for
degenerate states, especially in selecting the model space~\cite{LindgrenBook:82}.
The structure of terms of the PT series often leads to one-
and two-particle operators which almost in all versions of many-body perturbation 
theory are not in irreducible tensorial form and which can not use advantage of Racah algebra~\cite{Gaigalas_1996,Gaigalas_1997}.

Probably the most efficient and consistent way to account for  correlation and relativistic effects simultaneously
is to combine RCI and relativistic many-body perturbation theory methods~\cite{Bogetal:1997,DzuFla:07}. This is particularly relevant 
for complex many-electron atoms with open $f$-shells, such as lanthanides and actinides, 
when calculating energy spectra and other properties.

A combination of the relativistic configuration interaction method 
and the stationary second-order Rayleigh-Schr\"odinger many-body perturbation theory 
in an irreducible tensorial form~\cite{Gaigetal:2024} 
has been already developed and implemented in the {\sc Grasp}2018 package \cite{grasp2018}
to account for the core-valence correlations.
In this paper we present the extension of this method (see Section~\ref{Sec:Implementation}) 
when core correlations are added,
and show how to use it in real applications (see Section~\ref{Sec:Calculations}).

\section{Combination of RCI method
with the stationary second-order Rayleigh-Schr\"odinger many-body perturbation theory}
\label{Sec:Implementation}

In regular {\sc Grasp} \cite{grasp2018,grasp2013} calculations, to obtain the radial orbitals, the main correlations are included 
in the MCDHF calculations. 
This is followed by RCI calculations to add the other important correlations via the configuration state functions (CSFs).
To include valence (V) and valence-valence (VV) correlations,
the space is extended by the CSFs, 
in which one or two electrons are excited from valence orbitals to virtual orbitals. 
Other types of correlations, such as core-valence (CV) and core (C) correlations, are added in the same manner.

In this paper, we present a combination of the RCI method 
 and the stationary second-order Rayleigh-Schr\"odinger many-body perturbation theory 
in an irreducible tensorial form (RCI (RSMBPT)),
where CV and C correlations are included according to the RSMBPT. The inclusion of CV correlations using the RCI (RSMBPT) method has already been developed, 
and can be found in Ref. \cite{Gaigetal:2024} with the expressions and its implementation in the {\sc Grasp} package.
This contribution coming from the CV correlations of the configurations $K'$ to $E (K \chi J)$ in the second-order of perturbation theory is expressed as (see \cite[Eq. (22)]{Gaigetal:2024})
\begin{eqnarray}
\label{eq:BogEnergy_PT_CV}
  \Delta E_{PT(CV)} =
	\nonumber \\  [0.2cm]
& & \hspace*{-2.5cm}
   =  \Delta \mathcal{E}_0 \left(K J \right) +
	\nonumber \\  [0.2cm]
& & \hspace*{-2.5cm}
	+ \; \sum_{n \ell j} \sum_{k>0} \widetilde{f}_k \left( \ell j^{w}, \; K \chi J  \right)
	\Delta \mathcal{F}^{k} \left( n \ell j, \; n \ell j \right) +
	\nonumber \\
& & \hspace*{-2.5cm}
	+ \; \sum_{n \ell j} \sum_{n' \ell'j' > n \ell j} \left\{ \sum_{k>0} \widetilde{f}_k \left( \ell j^{w} \; \ell' j'^{w'},
	\; K \chi J  \right) \right. \times
	\nonumber \\ [0.2cm]
& & 
\times
 \Delta \mathcal{F}^{k} \left( n \ell j, \; n' \ell' j' \right) +
	\nonumber \\ [0.2cm]
& &  \hspace*{-2.5cm}
	+ \sum_{k} \widetilde{g}_k \left( \ell j^{w} \; \ell' j'^{w'}, \; K \chi J  \right)	
 \Delta \mathcal{G}^{k} \left( n \ell j, \; n' \ell' j' \right) +
	\nonumber \\ [0.2cm]
& & \hspace*{-2.5cm}
	+ \sum_{k} \widetilde{v}_k \left( \ell j^{w} \; \ell' j'^{w'}, \ell j^{w-2} \; \ell' j'^{w'+2},
	\; K \chi J \; K' \chi' J \right)	\times
\nonumber \\ [0.2cm]
& & 
\Bigg. \times
\Delta \mathcal{R}^{k} \left( n \ell j n \ell j, \; n' \ell' j' n '\ell' j' \right) \Bigg\} .
\end{eqnarray}

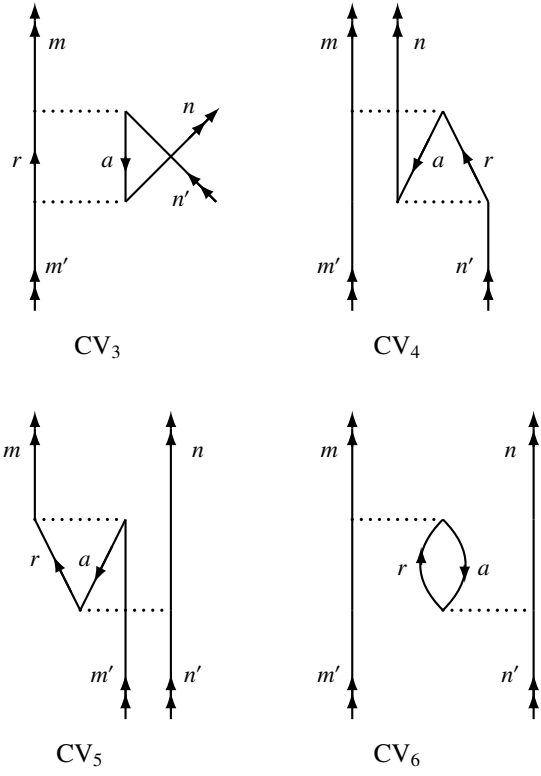
\begin{figure}
\begin{center}
\setlength{\unitlength}{1.2mm}
\begin{picture}(68,88)
\thicklines
\put(10,75){\line(0,1){10}}
\put(10,83){\vector(0,1){2}}
\put(10,85){\vector(0,1){2}}
\put(12.5,83){\makebox(0,0)[t]{\small{$m$}}}
\put(20,75){\line(1,-1){10}}
\put(28,73){\vector(1,1){2.4}}
\put(27,72){\vector(1,1){2}}
\put(27,76){\makebox(0,0)[t]{\small{$n$}}}
\multiput(10,75)(1,0){10}{\circle*{0.35}}
\put(10,65){\line(0,10){10}}
\put(10,69){\vector(0,1){2}}
\put(8,70){\makebox(0,0)[t]{\small{$r$}}}
\put(20,65){\line(0,10){10}}
\put(20,70){\vector(0,-1){2}}
\put(18,70){\makebox(0,0)[t]{\small{$a$}}}
\multiput(10,65)(1,0){10}{\circle*{0.35}}
\put(10,55){\line(0,1){10}}
\put(10,55){\vector(0,1){3}}
\put(12.5,58){\makebox(0,0){\small{$m^{\prime}$}}}
\put(10,53){\vector(0,1){3}}
\put(20,65){\line(1,1){10}}
\put(29,66){\vector(-1,1){2.4}}
\put(26.2,65.4){\makebox(0,0){\small{$n'$}}}
\put(30,65){\vector(-1,1){2}}
\put(17,49){\makebox(0,0){$\text{CV}_{3}$}}
\put(45,75){\line(0,1){10}}
\put(45,83){\vector(0,1){2}}
\put(45,85){\vector(0,1){2}}
\put(42.5,83){\makebox(0,0)[t]{\small{$m$}}}
\put(55,75){\line(0.5,-1){5}}
\put(52.5,83){\makebox(0,0)[t]{\small{$n$}}}
\multiput(45,75)(1,0){10}{\circle*{0.35}}
\put(45,65){\line(0,10){10}}
\put(60,70){\makebox(0,0)[t]{\small{$r$}}}
\put(50,83){\vector(0,1){2}}
\put(50,85){\vector(0,1){2}}
\put(50,65){\line(0,10){20}}
\put(53.5,72){\vector(-0.5,-1){2}}
\put(54.5,70){\makebox(0,0)[t]{\small{$a$}}}
\multiput(50,65)(1,0){10}{\circle*{0.35}}
\put(45,55){\line(0,1){10}}
\put(45,55){\vector(0,1){3}}
\put(42.5,58){\makebox(0,0){\small{$m^{\prime}$}}}
\put(45,53){\vector(0,1){3}}
\put(50,65){\line(0.5,1){5}}
\put(60,55){\line(0,1){10}}
\put(60,55){\vector(0,1){3}}
\put(60,53){\vector(0,1){3}}
\put(57.5,58){\makebox(0,0){\small{$n^{\prime}$}}}
\put(59,67){\vector(-0.5,1){2}}
\put(50,49){\makebox(0,0){$\text{CV}_{4}$}}
\put(10,30){\line(0,1){10}}
\put(10,38){\vector(0,1){2}}
\put(10,40){\vector(0,1){2}}
\put(7.5,38){\makebox(0,0)[t]{\small{$m$}}}
\put(10,30){\line(0.5,-1){5}}
\put(28,38){\makebox(0,0)[t]{\small{$n$}}}
\multiput(10,30)(1,0){10}{\circle*{0.35}}
\put(20,20){\line(0,10){10}}
\put(10,26){\makebox(0,0)[t]{\small{$r$}}}
\put(25,38){\vector(0,1){2}}
\put(25,40){\vector(0,1){2}}
\put(25,20){\line(0,10){20}}
\put(18.5,27){\vector(-0.5,-1){2}}
\put(15.5,26){\makebox(0,0)[t]{\small{$a$}}}
\multiput(15,20)(1,0){10}{\circle*{0.35}}
\put(20,10){\line(0,1){10}}
\put(20,10){\vector(0,1){3}}
\put(17.5,13){\makebox(0,0){\small{$m^{\prime}$}}}
\put(20,08){\vector(0,1){3}}
\put(15,20){\line(0.5,1){5}}
\put(25,10){\line(0,1){10}}
\put(25,10){\vector(0,1){3}}
\put(25,08){\vector(0,1){3}}
\put(27.5,13){\makebox(0,0){\small{$n^{\prime}$}}}
\put(14,22){\vector(-0.5,1){2}}
\put(15,04){\makebox(0,0){$\text{CV}_{5}$}}
\put(45,30){\line(0,1){10}}
\put(45,38){\vector(0,1){2}}
\put(45,40){\vector(0,1){2}}
\put(42.5,38){\makebox(0,0)[t]{\small{$m$}}}
\multiput(45,30)(1,0){10}{\circle*{0.35}}
\put(45,10){\line(0,1){10}}
\put(45,10){\vector(0,1){3}}
\put(42.5,13){\makebox(0,0){\small{$m^{\prime}$}}}
\put(45,08){\vector(0,1){3}}
\put(45,20){\line(0,10){10}}
\qbezier(55,20)(50,25)(55,30)
\put(52.6,25){\vector(0,1){2}}
\put(50.5,25){\makebox(0,0)[t]{\small{$r$}}}
\qbezier(55,20)(60,25)(55,30)
\put(57.4,25){\vector(0,-1){2}}
\put(59.5,25){\makebox(0,0)[t]{\small{$a$}}}
\multiput(55,20)(1,0){10}{\circle*{0.35}}
\put(65,30){\line(0,1){10}}
\put(65,38){\vector(0,1){2}}
\put(65,40){\vector(0,1){2}}
\put(62.5,38){\makebox(0,0)[t]{\small{$n$}}}
\put(65,20){\line(0,10){10}}
\put(65,08){\vector(0,1){3}}
\put(65,10){\line(0,1){10}}
\put(65,10){\vector(0,1){3}}
\put(62.5,13){\makebox(0,0){\small{$n^{\prime}$}}}
\put(50,04){\makebox(0,0){$\text{CV}_{6}$}}
\end{picture}
\caption{The CV Feynman diagrams of the second-order effective Hamiltonian for core correlations 
$(n_{a} \ell_{a})\, j_{a}^{2j_a+1}\, (n_{m} \ell_{m})\, j_{m}^{w_m} \,
\rightarrow \, (n_{a} \ell_{a})\, j_{a}^{2j_a} \, (n_{m} \ell_{m})\, j_{m}^{w_m} \, (n_{r} \ell_{r})\, j_{r}$.}
\label{Feynman}
\end{center}
\end{figure}

Here we extend the combination of RCI method
with the RSMBPT to include the core correlations via perturbation theory, which has already been used for CV correlations in \cite{Gaigetal:2024}. This allows even greater reduction of the CSF space for RCI method. The Feynman diagrams corresponding to the following type of core correlation are shown in Fig. \ref{Feynman}
where all lines with double arrow of diagrams are renamed $m$, i.e. $m' \equiv m$
\begin{equation}
\label{eq:C_cor}
    (n_{a} \ell_{a})\, j_{a}^{2j_a+1} \, (n_{m} \ell_{m})\, j_{m}^{w_m}  
   \rightarrow (n_{a} \ell_{a})\, j_{a}^{2j_a} \, (n_{m} \ell_{m})\, j_{m}^{w_m} \, (n_{r} \ell_{r})\, j_{r}.
\end{equation}
These are the same four two-particle Feynman diagrams which partly describe CV correlations (see \cite{Gaigetal:2024}). The detailed explanations and expressions in irreducible tensorial form of the diagrams are presented in Ref. \cite{Gaigetal:2024}.

It should also be mentioned that the method proposed in the paper does not allow the following type of core and core-valence correlations to be included via RSMBPT
\begin{equation}
\label{eq:not-C-a}
    (n_{a} \ell_{a})\, j_{a}^{2j_a+1} \, (n_{m} \ell_{m})\, j_{m}^{w_m} 
   \rightarrow (n_{a} \ell_{a})\, j_{a}^{2j_a} \; (n_{m} \ell_{m})\, j_{m}^{w_m+1} 
\end{equation}
\begin{eqnarray}
\label{eq:not-CV-a}
    (n_{a} \ell_{a})\, j_{a}^{2j_a+1} \, (n_{m} \ell_{m})\, j_{m}^{w_m} \, (n_{n} \ell_{n})\, j_{n}^{w_n}  \rightarrow 
	\nonumber \\ [0.2cm]
& & \hspace*{-3.0cm}
   \rightarrow (n_{a} \ell_{a})\, j_{a}^{2j_a} \; (n_{m} \ell_{m})\, j_{m}^{w_m-1} \, (n_{n} \ell_{n})\, j_{n}^{w_n+2}. 
\end{eqnarray}
This is related to the fact that these C (Eq. (\ref{eq:not-C-a})) and CV (Eq. (\ref{eq:not-CV-a})) correlations described via
Feynman diagrams depend on the potential of the radial orbitals found. This makes the implementation of this type of correlations impossible to be included in the RSMBPT method for any potential of MCDHF equations.  
Therefore, these correlations should be included in the RCI calculations in a regular way.


According to \cite{Bogetal:1997}, the energy contribution of the admixed configurations from C correlations (Eq. (\ref{eq:C_cor})) can be added to a regular energy of the 
term $\chi J$ of the configuration $K$ and can be
expressed as the energy $E_0 \left(K J \right)$, which does not depend on the term, and the sum of product of Slater integrals and spin-angular coefficients
\begin{eqnarray}
\label{eq:BogEnergy}
   E\left(K \chi J \right)
   = E_0 \left(K J \right) + \Delta \mathcal{E}_0 \left(K J \right) +
	\nonumber \\  [0.2cm]
& & \hspace*{-3.0cm}
	+ \; \sum_{n \ell j} \sum_{k>0} \widetilde{f}_k \left( \ell j^{w}, \; K \chi J  \right)
	\left[ F^{k} \left( n \ell j, \; n \ell j \right)  + \right.
	\nonumber \\  [0.2cm]
& &  
\left.	+ \Delta \mathcal{F}^{k} \left( n \ell j, \; n \ell j \right) \right] .
\end{eqnarray}
The contribution of the C correlations in the second-order of perturbation theory according to Eq. (\ref{eq:BogEnergy}) can be expressed as 
\begin{eqnarray}
\label{eq:BogEnergy_PT_C}
  \Delta E_{PT(C)}
   =  \Delta \mathcal{E}_0 \left(K J \right) +
	\nonumber \\  [0.2cm]
& & \hspace*{-1.5cm}
	+ \; \sum_{n \ell j} \sum_{k>0} \widetilde{f}_k \left( \ell j^{w}, \; K \chi J  \right)
	\Delta \mathcal{F}^{k} \left( n \ell j, \; n \ell j \right) .
\end{eqnarray}

\begin{table*}
\begin{center}
\begin{tabular}{|l|} \hline
$\Delta \mathcal{E}_0$ corrections \\ \hline \hline
\\
$\overbrace{(n_{a} \ell_{a})\, j_{a}^{2j_a+1}}^{\text{core subshells}} \, \overbrace{(n_{m} \ell_{m})\, j_{m}^{w_m}}^{\text{valence subshells}} \;
\rightarrow \; \overbrace{(n_{a} \ell_{a})\, j_{a}^{2j_a}}^{\text{core subshells}} \; \overbrace{(n_{m} \ell_{m})\, j_{m}^{w_m}}^{\text{valence subshells}} \; \overbrace{(n_{r} \ell_{r})\, j_{r}}^{\text{virtual subshells}}$ \\
\\
$\frac{w_m }{\left[j_m\right]} \left\{ \underbrace{ \left( w_m - 1 - \left[ j_m\right] \right)
 \,
\mathcal{A}\left( 0, \, m a, \, r m \right)}_{\text{from $\text{CV}_{3}$ Feynman diagram}}
\underbrace{- \; \left(w_m - 1 \right) \;
\mathcal{D}(ma,mr) \; \delta \left(j_r,j_{a}\right)}_{\text{from $\text{CV}_{4},\text{CV}_{5}$ and $\text{CV}_{6},$ Feynman diagrams}}  \right\}$\\
\\
$+ \frac{w_m (-1)^{j_a+j_r}}{\left[j_m\right]}  \left\{ \underbrace{2 \sum_{k} \mathcal{C}\left( k, \, m a, \, m r \right)}_{\text{from $\text{CV}_{4}\text{ and CV}_{5}$ Feynman diagrams}}
+ \underbrace{\sum_{k} \frac{1}{\left[ k \right]} \mathcal{P} \left( kk, ma, \, mr \right)}_{\text{from $\text{CV}_{6}$ Feynman diagram}} \right\}$ \\
\\ \hline 
\end{tabular}
\end{center}
\caption{Expressions for core corrections to the energy in Eq. (\ref{eq:BogEnergy}), not depending on term.}
\label{tab:Implemen_C1}
\end{table*}

$\Delta \mathcal{E}_0 \left(K J \right)$ (see Table~\ref{tab:Implemen_C1}) - the contribution of C correlations in the second-order of perturbation theory deriving from the total energy $E_0 \left(K J \right)$ can be expressed through $\mathcal{A}$, $\mathcal{D}$ and $\mathcal{C}$ coefficients, which have the following expressions
 
\begin{eqnarray}
\label{eq:BogA}
   \mathcal{A}\left(x, \; i j, \; i' j'\right) =
	\nonumber \\  [0.2cm]
& & \hspace*{-1.5cm}
  = \sum_{k,k'}
		  \left\{
    \begin{array}{ccc}
      k  & k' & x \\
      j_{i} & j_{i} & j_{i'}
    \end{array} \right\}
				  \left\{
    \begin{array}{ccc}
      k  & k' & x \\
      j_{j'} & j_{j'} & j_{j}
    \end{array} \right\}
\mathcal{P}\left(kk', \; i j, \; i' j'\right) ,
\end{eqnarray}

\begin{eqnarray}
\label{eq:BogD}
   \mathcal{D}\left( i j, \; i' j'\right) 
  = 
\mathcal{P}\left(00, \; i j, \; i' j'\right) +
	\nonumber \\  [0.2cm]
& & \hspace*{-2.0cm}
+ \frac{2}{\sqrt{\left[ j_i, j_j \right]}}
\sum_{k}
\left( -1 \right)^{j_i+j_j+k}
\mathcal{Q}\left(0k, \; i j, \; i' j'\right),
\end{eqnarray}

\begin{equation}
\label{eq:BogC}
   \mathcal{C}\left(k, \; i j, \; i' j'\right) 
  = \sum_{k'}
		  \left\{
    \begin{array}{ccc}
      k  & j_{i} & j_{i'} \\
      k' & j_{j} & j_{j'}
    \end{array} \right\}
\mathcal{Q}\left(kk', \; i j, \; i' j'\right) ,
\end{equation}

\begin{equation}
\label{eq:BogP}
   \mathcal{P}\left(kk', \; i j, \; i' j'\right) 
   = \mathcal{R}^{k}\left(i j, \; i' j'\right) \; \mathcal{R}^{k'}\left(i' j', \; i j \right)  \; 
	\mathcal{O}\left(K', K \right),
\end{equation}

\begin{equation}
\label{eq:BogQ}
   \mathcal{Q}\left(kk', \; i j, \; i' j'\right)
   = \mathcal{R}^{k}\left(i j, \; i' j'\right) \; \mathcal{R}^{k'}\left(i' j', \; j i\right)  \; 
\mathcal{O}\left(K', K \right) .
\end{equation}
We would like to emphasize that the energy denominator is defined differently (with the opposite sign) than in the expressions of Feynman diagrams (see for example~\cite[Fig.~3]{Gaigetal:2024}, Eqs. (\ref{eq:BogP}) and (\ref{eq:BogQ})).

\begin{equation}
\label{eq:BogO1}
\mathcal{O}\left(K', K \right)
= \frac{1}{\overline{E}\left(K' \right)-\overline{E}\left(K\right)}
\end{equation}

The $\mathcal{R}^{k}$ is the generalized integral of electrostatic interaction between electrons in 
the Table~\ref{tab:Implemen_C1}
\begin{eqnarray}
\label{eq:BogRk}
   \mathcal{R}^{k}\left(i j, i' j'\right)
   = \bigg\{ \Big[ 1 + \delta \left( i, j \right) \Big]  \Big[ 1 + \delta \left( i', j' \right) \Big]  \bigg\} ^{-1/2} \times
	\nonumber \\  [0.2cm]
& & \hspace*{-4.5cm}
	 \times \, R^{k}\left(n_i j_i \, n_jj_j, \, n_{i'}j_{i'} \, n_{j'}j_{j'} \right) \times
	\nonumber \\  [0.2cm]
& & \hspace*{-4.5cm}
	   \times \redmem{\ell_i j_{i}}{\, C^{(k)} \,}{ \ell_{i'} j_{i'}}
     \redmem{\ell_j j_{j}}{\, C^{(k)} \,}{ \ell_{j'} j_{j'}}, 
\end{eqnarray}
where $R^{k}\left(n_i j_i \, n_jj_j, \, n_{i'}j_{i'} \, n_{j'}j_{j'} \right)$ is the radial integral.

\begin{table*}
\begin{center}
\begin{tabular}{|lll|} \hline
Corrections & Slater integral & $k$ values\\ \hline  \hline
& & \\
$\underbrace{ 
2 \left[k \right] \mathcal{Y}\left( 1, \, k, \, m a, \, r m, \, m m \right)}_{\text{from $\text{CV}_3$ Feynman diagram}}$ 
&$\Delta \mathcal{F}^{k}(m,m)$ & $k>0$\\
&&\\
$+ \left( -1 \right)^{j_a+j_r+k} \left\{   
\underbrace{4 \, \mathcal{Z}\left(1, k, \, m a, \, m r, \; m m \right) }_{\text{from $\text{CV}_4$ and $\text{CV}_5$ Feynman diagrams}} + \underbrace{\frac{2}{\left[ k \right] } \mathcal{P}\left( kk, \, m a, \, m r  \right)}_{\text{from $\text{CV}_6$ Feynman diagram}} \right\}  $ 
& & \\ \hline 
\end{tabular}
\end{center}
\caption{Expressions for Slater integrals $\Delta \mathcal{F}^{k}(m,m)$ (see Eq. (\ref{eq:BogEnergy})) corresponding the core 
$(n_{a} \ell_{a})\, j_{a}^{2j_a+1} \, (n_{m} \ell_{m})\, j_{m}^{w_m} 
   \rightarrow (n_{a} \ell_{a})\, j_{a}^{2j_a} \; (n_{m} \ell_{m})\, j_{m}^{w_m} \; (n_{r} \ell_{r})\, j_{r}$  correlations.} 
\label{tab:Implemen_C2}
\end{table*}

The contribution of C correlations in the second-order of perturbation theory arriving at $F^{k} \left( n \ell j, \; n \ell j \right)$, $\Delta \mathcal{F}^{k} \left( n \ell j, \; n \ell j \right)$ (see Table~\ref{tab:Implemen_C2}) can be expressed through $\mathcal{Y}$ and $\mathcal{Z}$ coefficients. These have the following expressions

\begin{eqnarray}
\label{eq:BogY}
   \mathcal{Y}\left(1, \; x, \; i j, \; i' j', \; i'' j''\right) =
	\nonumber \\  
& & \hspace*{-2.5cm}
  = \sum_{k,k'}
		\left( -1 \right)^{k+k'+x}
				  \left\{
    \begin{array}{ccc}
      k  & k' & x \\
      j_{i''} & j_{i} & j_{i'}
    \end{array} \right\}				  \left\{
    \begin{array}{ccc}
      k  & k' & x \\
      j_{j''} & j_{j'} & j_{j}
    \end{array} \right\} \times
\nonumber \\ 
& & 
\times \, 
\mathcal{T}\left(1, \; kk', \; i j, \; i' j', \; i'' j''\right) ,
\end{eqnarray}

\begin{eqnarray}
\label{eq:BogT}
   \mathcal{T}\left(1, \; kk', \; i j, \; i' j', \; i'' j''\right) = 
	\nonumber \\  
& & \hspace*{-2.0cm}
  =
\mathcal{R}^{k}\left(i j, \; i' j'\right)  \; 
\mathcal{R}^{k'}\left(i' j'', \; i'' j \right)  \; 
\mathcal{O}\left(K', K\right) ,
\end{eqnarray}

\begin{eqnarray}
\label{eq:BogZ}
   \mathcal{Z}\left(1, \; k, \; i j, \; i' j', \; i'' j''\right) =
	\nonumber \\  
& & \hspace*{-2.5cm}
  = \sum_{k'}
				  \left\{
    \begin{array}{ccc}
      k & j_{i''} & j_{j''} \\
      k' & j_{j} & j_{j'}
    \end{array} \right\}
\mathcal{U}\left(1, \; kk', \; i j, \; i' j', \; i'' j''\right) ,
\end{eqnarray}
\begin{eqnarray}
\label{eq:BogU}
   \mathcal{U}\left(1, \; kk', \; i j, \; i' j', \; i'' j''\right) = 
	\nonumber \\  
& & \hspace*{-2.0cm}
	= 
   \mathcal{R}^{k}\left(i j, \; i' j'\right)  \; 
\mathcal{R}^{k'}\left(j'' j', \; j i'' \right)  \; 
\mathcal{O}\left(K', K\right) .
\end{eqnarray}

\section{Calculation of core-valence and core correlations with a new approach}
\label{Sec:Calculations}

As described in the above section, the method
based on the Rayleigh-Schr\"odinger perturbation theory in an irreducible tensorial form \cite{Gaigetal:2024}
is extended to include C correlations in the computations. 
This section aims to present the results when the CV and C correlations are 
included in regular way and using the stationary second-order Rayleigh-Schr\"odinger many-body
perturbation theory in an irreducible tensorial form. 
For this purpose we computed the energy levels of the $\mathrm{3s^2}$, $\mathrm{3p^2}$, $\mathrm{3s3d}$, $\mathrm{3d^2}$, 
$\mathrm{3p3d}$, $\mathrm{3s3p}$ configurations of the Fe XV.
Below the computational procedure and results from regular {\sc Grasp}2018 and RCI (RSMBPT) computations are presented.

\subsection{Computational scheme}
An initial MCDHF calculation for 
even and odd states of the $\mathrm{3s^2}$, $\mathrm{3p^2}$, $\mathrm{3s3d}$, $\mathrm{3d^2}$, 
$\mathrm{3p3d}$, $\mathrm{3s3p}$ 
configurations was done in the extended optimal level (EOL) scheme \citep{EOL}.
The initial calculation was followed by
separate calculations in the EOL scheme for the even and odd parity states. 
The space of CSFs, referred to as the active space (AS), building the atomic state function (ASFs) 
was obtained using the multireference-single-double (MR-SD) method \citep{Fisetal:16a}.
The MR set consists of the $\mathrm{3s^2}$, $\mathrm{3p^2}$, $\mathrm{3s3d}$, $\mathrm{3d^2}$ even and 
$\mathrm{3p3d}$, $\mathrm{3s3p}$ odd configurations. The 
orbital spaces (OS), to which single and double (SD) substitutions from the configurations in the MR were allowed, 
are: $OS_1$ = \{4s, $\mathrm{4p_-}$, 4p, $\mathrm{4d_-}$, 4d, $\mathrm{4f_-}$, 4f\}, ..., 
$OS_5$ = \{8s, $\mathrm{8p_-}$, 8p, $\mathrm{8d_-}$, 8d, $\mathrm{8f_-}$, 8f, $\mathrm{8g_-}$, 8g, $\mathrm{8i_-}$, 8i\}.
SD substitutions were allowed from the 3s, $\mathrm{3p_-}$, 3p, $\mathrm{3d_-}$, 3d orbitals, 
and S substitutions were allowed only from the 2s or $\mathrm{2p_-}$ and 2p core orbital.
Only CSFs that have non-zero
matrix elements with the CSFs belonging to the configurations
in the MR were retained.
No substitutions were allowed from the $\mathrm{1s^2}$ core, which defines an inactive closed core.
Based on the orbitals from the MCDHF calculations, 
RCI calculations were further performed, including the Breit interaction and 
leading quantum electrodynamic (QED) effects -- the vacuum polarization and the self-energy corrections.
Regular RCI calculations 
are marked as \textbf{CV+C RCI}.

The results of the calculation when the CV and C correlations are included according to the RSMBPT method are marked as \textbf{CV+C RCI (RSMBPT)}. 
In this case, we use a program to determine the contribution 
of each $K'$ configuration of the CV and C correlations for CSF for which energy needs to be calculated according to 
Rayleigh-Schr\"odinger perturbation theory in an irreducible tensorial form according to the Eqs. (\ref{eq:BogEnergy_PT_CV}) and (\ref{eq:BogEnergy_PT_C}).
The program calculates the total contribution of the CV and C correlations  
and the contribution of each $K'$ configuration of these correlations for the computed levels.
$K'$ configurations are sorted in descending order according to the impact of the CV and C correlations for each level.
Further, we select $K'$ configurations by the CV and C correlations impact
with the specified fraction (expressed in the percentage) of the total CV and C contribution, and perform RCI computations including them.
The calculations using the RSMBPT method were carried out 
using different amounts of CV and C correlations: 95\%, 99\%, 99.5\%, 99.95\%, and 100\%.
In the RSMBPT method the CSF space is divided into three sets: $F$, $F'$ and $G$ 
(see Ref. \cite{Gaigetal:2024} in details).
Thus in RSMBPT computations, the 1s is also defined as inactive core subshell, 2s, $\mathrm{2p_-}$ and 2p subshells are defined as active core subshells 
(that correspond to $F$ set), 3s, $\mathrm{3p_-}$, 3p, $\mathrm{3d_-}$, and 3d as valence subshells (that correspond to $F'$ set), and subshells belonging to $OS_1$, ..., $OS_5$ as virtual ones (that correspond to $G$ set). This distribution of space is consistent with regular {\sc Grasp} calculations
and allows the use of a combination of RCI and RSMBPT methods.
It should be noted that the program gives the contribution of the CV and C correlations of $K'$ configuration with the value
greater than {\tt 1.0E-11}, the rest contributions are neglected. 
Therefore a number of CSFs, when 100\% of CV and C correlations are included, is smaller than in the regular \textbf{CV+C RCI} calculations.
The C (Eq. (\ref{eq:not-C-a})) and CV (Eq. (\ref{eq:not-CV-a})) correlations (mentioned above in Section \ref{Sec:Implementation}) which were not included with RSMBPT method, were added to RCI calculations in a regular way, together with the valence and valence-valence correlations.

\subsection{Results}

\begin{table*}[!ht]
{\scriptsize
\caption{The total energies (in a.u.) from \textbf{CV+C RCI} calculations and differences (in a.u.) between \textbf{CV+C RCI (RSMBPT)} and \textbf{CV+C RCI} 
 energies ($\Delta E_{\textbf{(CV+C~RCI (RSMBPT))-(CV+C~RCI)}}$) for Fe XV are given when CV and C correlations are included in the computations.}            
\label{comp1}
\centering
\begin{tabular}{r l r r r r r r}
\hline\hline
\multicolumn{1}{c}{\multirow{2}{*}{No.}} & \multicolumn{1}{c}{\multirow{2}{*}{State}} & \multicolumn{1}{c}{\multirow{2}{*}{\textbf{CV+C RCI}}}& \multicolumn{5}{c}{$\Delta E_{\textbf{(CV+C~RCI (RSMBPT))-(CV+C~RCI)}}$} \\
\cline{4-8}
&&& \multicolumn{1}{c}{95\%} & \multicolumn{1}{c}{99\%} & \multicolumn{1}{c}{99.5\%} & \multicolumn{1}{c}{99.95\%} & \multicolumn{1}{c}{100\%} \\
\hline
\noalign{\smallskip}
 1& $\mathrm{3s^2~^1S_0}$    &  -1182.43030203& 0.00178946& 0.00045030& 0.00026337& 0.00008561&  0.00000002  \\
 2& $\mathrm{3s3p~^3P^o_0}$  &  -1181.36480226& 0.00197381& 0.00058056& 0.00034857& 0.00008929&  0.00000039  \\ 
 3& $\mathrm{3s3p~^3P^o_1}$  &  -1181.33820345& 0.00160825& 0.00039368& 0.00022404& 0.00004079&  0.00000010  \\
 4& $\mathrm{3s3p~^3P^o_2}$  &  -1181.27370668& 0.00167068& 0.00041319& 0.00024809& 0.00004752&  0.00000002  \\
 5& $\mathrm{3s3p~^1P^o_1}$  &  -1180.82591412& 0.00154878& 0.00037164& 0.00021696& 0.00004166&  0.00000018  \\
 6& $\mathrm{3p^2~^3P_0}$    &  -1179.90299188& 0.00227387& 0.00058237& 0.00036949& 0.00012520&  0.00000003  \\ 
 7& $\mathrm{3p^2~^1D_2}$    &  -1179.87922968& 0.00184307& 0.00044139& 0.00023671& 0.00004951&  0.00000002  \\
 8& $\mathrm{3p^2~^3P_1}$    &  -1179.85723426& 0.00210413& 0.00054570& 0.00032064& 0.00008543&  0.00000004  \\
 9& $\mathrm{3p^2~^3P_2}$    &  -1179.77862047& 0.00159810& 0.00035493& 0.00019870& 0.00004506&$-$0.00000004 \\
10& $\mathrm{3p^2~^1S_0}$    &  -1179.42248968& 0.00220519& 0.00053086& 0.00031971& 0.00009746&  0.00000002  \\ 
11& $\mathrm{3s3d~^3D_1}$    &  -1179.33608105& 0.00198106& 0.00044895& 0.00026079& 0.00005123&$-$0.00000037 \\
12& $\mathrm{3s3d~^3D_2}$    &  -1179.33137782& 0.00161881& 0.00032508& 0.00016488& 0.00002351&  0.00000026  \\
13& $\mathrm{3s3d~^3D_3}$    &  -1179.32397202& 0.00201250& 0.00047993& 0.00023062& 0.00004272&$-$0.00000133 \\
14& $\mathrm{3s3d~^1D_2}$    &  -1178.95545428& 0.00162037& 0.00034608& 0.00017830& 0.00002806&$-$0.00000068 \\
15& $\mathrm{3p3d~^3F^o_2}$  &  -1178.19871370& 0.00202851& 0.00047531& 0.00027051& 0.00005367&  0.00000157  \\  
16& $\mathrm{3p3d~^3F^o_3}$  &  -1178.15359232& 0.00230291& 0.00055974& 0.00032274& 0.00007092&  0.00000069  \\
17& $\mathrm{3p3d~^1D^o_2}$  &  -1178.10669754& 0.00181324& 0.00040713& 0.00022404& 0.00005106&$-$0.00000019 \\
18& $\mathrm{3p3d~^3F^o_4}$  &  -1178.10109503& 0.00250480& 0.00062540& 0.00036157& 0.00009234&$-$0.00000070 \\
19& $\mathrm{3p3d~^3D^o_1}$  &  -1177.95043389& 0.00197565& 0.00045613& 0.00025318& 0.00005326&  0.00000093  \\ 
20& $\mathrm{3p3d~^3P^o_2}$  &  -1177.94726099& 0.00188294& 0.00042846& 0.00023824& 0.00005289&$-$0.00000069 \\
21& $\mathrm{3p3d~^3D^o_3}$  &  -1177.89568608& 0.00217807& 0.00052526& 0.00028042& 0.00006212&$-$0.00000270 \\
22& $\mathrm{3p3d~^3P^o_0}$  &  -1177.89056768& 0.00260665& 0.00064465& 0.00040704& 0.00009653&  0.00000117  \\ 
23& $\mathrm{3p3d~^3P^o_1}$  &  -1177.88908357& 0.00176233& 0.00040870& 0.00021420& 0.00004491&  0.00000106  \\
24& $\mathrm{3p3d~^3D^o_2}$  &  -1177.88749171& 0.00170255& 0.00038866& 0.00020909& 0.00004652&  0.00000008  \\
25& $\mathrm{3p3d~^1F^o_3}$  &  -1177.58555544& 0.00222814& 0.00055158& 0.00030917& 0.00006694&  0.00000025  \\
26& $\mathrm{3p3d~^1P^o_1}$  &  -1177.52845259& 0.00194221& 0.00043948& 0.00024891& 0.00005826&  0.00000100  \\
27& $\mathrm{3d^2~^3F_2}$    &  -1176.18298893& 0.00152815& 0.00030221& 0.00014706& 0.00002265&  0.00000108  \\
28& $\mathrm{3d^2~^3F_3}$    &  -1176.17531388& 0.00218646& 0.00052794& 0.00027605& 0.00005240&$-$0.00000086 \\
29& $\mathrm{3d^2~^3F_4}$    &  -1176.16597541& 0.00203061& 0.00043858& 0.00022350& 0.00004804&$-$0.00000154 \\
30& $\mathrm{3d^2~^1D_2}$    &  -1176.03470535& 0.00151666& 0.00029751& 0.00014078& 0.00002390&$-$0.00000021 \\
31& $\mathrm{3d^2~^3P_0}$    &  -1176.02162854& 0.00238518& 0.00056882& 0.00031729& 0.00007873&  0.00000002  \\
32& $\mathrm{3d^2~^3P_1}$    &  -1176.01869277& 0.00211055& 0.00048229& 0.00026501& 0.00006066&  0.00000001  \\
33& $\mathrm{3d^2~^1G_4}$    &  -1176.01407157& 0.00223271& 0.00050833& 0.00027154& 0.00005889&$-$0.00000401 \\
34& $\mathrm{3d^2~^3P_2}$    &  -1176.01176978& 0.00148483& 0.00029431& 0.00013336& 0.00002842&$-$0.00000035 \\
35& $\mathrm{3d^2~^1S_0}$    &  -1175.64578266& 0.00234712& 0.00055670& 0.00031665& 0.00007741&  0.00000003  \\
\hline
\noalign{\smallskip}
\multicolumn{2}{r}{$N_{CSFs}$} &    372043&    90859& 182643&  218556& 300041&  360394  \\
\hline
\hline
\end{tabular}
}
\end{table*}

\begin{table*}[!ht]
{\scriptsize
\caption{The energy levels (in cm$^{-1}$) and differences (in cm$^{-1}$) between \textbf{CV+C RCI} and NIST
 energies ($\Delta E_{\textbf{(CV+C~RCI)-(NIST)}}$), and between \textbf{CV+C RCI (RSMBPT)} and \textbf{CV+C RCI} 
 energies ($\Delta E_{\textbf{(CV+C~RCI (RSMBPT))-(CV+C~RCI)}}$) for Fe XV are given when CV and C correlations are included in the computations.}            
\label{comp2}
\centering
\begin{tabular}{r l r r r r r r r r r r}
\hline\hline
\multicolumn{1}{c}{\multirow{3}{*}{No.}} & \multicolumn{1}{c}{\multirow{3}{*}{State}} & \multicolumn{2}{c}{Energies} && \multicolumn{7}{c}{Energy differences} \\
\cline{3-4} \cline{6-12}
 &  & \multicolumn{1}{c}{\multirow{2}{*}{NIST}}& \multicolumn{1}{c}{\multirow{2}{*}{\textbf{CV+C RCI}}}&& \multicolumn{1}{c}{\multirow{2}{*}{$\Delta E_{\textbf{(CV+C~RCI)-(NIST)}}$}} && \multicolumn{5}{c}{$\Delta E_{\textbf{(CV+C~RCI (RSMBPT))-(CV+C~RCI)}}$} \\
\cline{8-12}
&&&&&&& \multicolumn{1}{c}{95\%} & \multicolumn{1}{c}{99\%} & \multicolumn{1}{c}{99.5\%} & \multicolumn{1}{c}{99.95\%} & \multicolumn{1}{c}{100\%} \\
\hline
\noalign{\smallskip}
 1& $\mathrm{3s^2~^1S_0}$    &       0&       0.00&&        &&         &         &         &         &         \\
 2& $\mathrm{3s3p~^3P^o_0}$  &  233842&  233850.17&&8.17   &&    40.46&    28.59&    18.70&     0.81&    0.08 \\ 
 3& $\mathrm{3s3p~^3P^o_1}$  &  239660&  239687.93&&7.93  && $-$39.77& $-$12.42&  $-$8.63&  $-$9.83&    0.02 \\
 4& $\mathrm{3s3p~^3P^o_2}$  &  253820&  253843.34&&23.34  && $-$26.07&  $-$8.15&  $-$3.36&  $-$8.36&    0.00 \\
 5& $\mathrm{3s3p~^1P^o_1}$  &  351911&  352122.45&&211.45 && $-$52.83& $-$17.27& $-$10.19&  $-$9.65&    0.03 \\
 6& $\mathrm{3p^2~^3P_0}$    &  554524&  554680.46&&156.46 &&   106.32&    28.99&    23.29&     8.69&    0.01 \\ 
 7& $\mathrm{3p^2~^1D_2}$    &  559600&  559895.66&&295.66 &&    11.77&  $-$1.95&  $-$5.85&  $-$7.92&    0.00 \\
 8& $\mathrm{3p^2~^3P_1}$    &  564602&  564723.10&&121.10 &&    69.06&    20.94&    12.57&  $-$0.04&    0.00 \\
 9& $\mathrm{3p^2~^3P_2}$    &  581803&  581976.83&&173.83 && $-$42.00& $-$20.93& $-$14.19&  $-$8.90& $-$0.01 \\
10& $\mathrm{3p^2~^1S_0}$    &  659627&  660138.51&&511.51 &&    91.24&    17.68&    12.36&     2.60&    0.00 \\ 
11& $\mathrm{3s3d~^3D_1}$    &  678772&  679103.01&&331.01 &&    42.05&  $-$0.30&  $-$0.57&  $-$7.55& $-$0.09 \\
12& $\mathrm{3s3d~^3D_2}$    &  679785&  680135.25&&350.25 && $-$37.45& $-$27.48& $-$21.62& $-$13.63&    0.05 \\
13& $\mathrm{3s3d~^3D_3}$    &  681416&  681760.63&&344.63 &&    48.96&     6.51&  $-$7.18&  $-$9.41& $-$0.29 \\
14& $\mathrm{3s3d~^1D_2}$    &  762093&  762640.93&&547.93 && $-$37.11& $-$22.87& $-$18.67& $-$12.63& $-$0.15 \\
15& $\mathrm{3p3d~^3F^o_2}$  &  928241&  928726.29&&485.29 &&    52.46&     5.49&     1.57&  $-$7.01&    0.34 \\  
16& $\mathrm{3p3d~^3F^o_3}$  &  938126&  938629.29&&503.29 &&   112.69&    24.01&    13.03&  $-$3.23&    0.14 \\
17& $\mathrm{3p3d~^1D^o_2}$  &  948513&  948921.50&&408.50 &&     5.22&  $-$9.47&  $-$8.63&  $-$7.58& $-$0.05 \\
18& $\mathrm{3p3d~^3F^o_4}$  &  949658&  950151.11&&493.11 &&   157.00&    38.43&    21.55&     1.48& $-$0.16 \\
19& $\mathrm{3p3d~^3D^o_1}$  &  982868&  983217.41&&349.41 &&    40.86&     1.28&  $-$2.24&  $-$7.10&    0.20 \\ 
20& $\mathrm{3p3d~^3P^o_2}$  &  983514&  983913.78&&399.78 &&    20.52&  $-$4.79&  $-$5.52&  $-$7.18& $-$0.16 \\
21& $\mathrm{3p3d~^3D^o_3}$  &  994852&  995233.16&&381.16 &&    85.29&    16.45&     3.75&  $-$5.15& $-$0.59 \\
22& $\mathrm{3p3d~^3P^o_0}$  &  995889&  996356.52&&467.52 &&   179.35&    42.66&    31.53&     2.40&    0.26 \\ 
23& $\mathrm{3p3d~^3P^o_1}$  &  996243&  996682.25&&439.25 &&  $-$5.96&  $-$9.13& $-$10.80&  $-$8.93&    0.23 \\
24& $\mathrm{3p3d~^3D^o_2}$  &  996623&  997031.62&&408.62 && $-$19.07& $-$13.53& $-$11.91&  $-$8.58&    0.01 \\
25& $\mathrm{3p3d~^1F^o_3}$  & 1062515& 1063298.97&&783.97 &&    96.28&    22.23&    10.05&  $-$4.10&    0.05 \\
26& $\mathrm{3p3d~^1P^o_1}$  & 1074887& 1075831.60&&944.60 &&    33.52&  $-$2.38&  $-$3.18&  $-$6.00&    0.21 \\
27& $\mathrm{3d^2~^3F_2}$    & 1370331& 1371126.74&&795.74 && $-$57.35& $-$32.50& $-$25.53& $-$13.82&    0.23 \\
28& $\mathrm{3d^2~^3F_3}$    & 1372035& 1372811.22&&776.22 &&    87.13&    17.04&     2.78&  $-$7.29& $-$0.19 \\
29& $\mathrm{3d^2~^3F_4}$    & 1374056& 1374860.78&&804.78 &&    52.92&  $-$2.58&  $-$8.75&  $-$8.25& $-$0.35 \\
30& $\mathrm{3d^2~^1D_2}$    & 1402592& 1403671.22&&1079.22&& $-$59.87& $-$33.53& $-$26.90& $-$13.54& $-$0.05 \\
31& $\mathrm{3d^2~^3P_0}$    &        & 1406541.25&&       &&   130.75&    26.01&    11.84&  $-$1.51&    0.00 \\
32& $\mathrm{3d^2~^3P_1}$    &        & 1407185.58&&       &&    70.47&     7.02&     0.36&  $-$5.48&    0.00 \\
33& $\mathrm{3d^2~^1G_4}$    & 1407058& 1408199.81&&1141.81&&    97.29&    12.74&     1.80&  $-$5.86& $-$0.88 \\
34& $\mathrm{3d^2~^3P_2}$    & 1407773& 1408705.00&&932.00 && $-$66.86& $-$34.24& $-$28.53& $-$12.55& $-$0.08 \\
35& $\mathrm{3d^2~^1S_0}$    & 1487054& 1489029.89&&1975.89&&   122.39&    23.35&    11.69&  $-$1.80&    0.00 \\
\hline                                                    
\noalign{\smallskip}
\multicolumn{2}{r}{$N_{CSFs}$} &&    \multicolumn{1}{c}{372043}&&&&    90859& 182643&  218556& 300041&  360394  \\
\hline
\hline
\end{tabular}
}
\end{table*}

Table \ref{comp1} presents the total energies from regular {\sc Grasp}2018 calculations (\textbf{CV+C RCI}) for 35 computed states and 
differences of energies using \textbf{CV+C RCI (RSMBPT)} method with \textbf{CV+C RCI} results ($\Delta E_{\textbf{(CV+C~RCI (RSMBPT))-(CV+C~RCI)}}$). 
In the last line of the table the number of CSFs ($N_{CSFs}$) from each calculation is given.
It is seen from the table that the results obtained with the RSMBPT method converge to regular RCI results (\textbf{CV+C RCI}) 
by including step by step the most important $K'$ configurations of CV and C correlations.
The calculations using the RSMBPT method are carried out in five steps, including 
95\%, 99\%, 99.5\%, 99.95\%, and 100\% of CV and C correlations. 
In case when all contributions of CV and C correlations (column 100\%) given by the program are included in the computations, 
the \textbf{CV+C RCI (RSMBPT)} reproduce the results of regular RCI computations. 
The negligible difference (till 4.0E-06 a.u.) in the 100\% column could be due to
omitted CV and C correlations with a very small contribution ({\tt 1.0E-11}) as it was mentioned above.

In Table \ref{comp2} the energy levels results from regular {\sc Grasp}2018 calculations \textbf{CV+C RCI} and 
from calculations using the RSMBPT method \textbf{CV+C RCI (RSMBPT)} are compared. 
There are also given energy levels from the Atomic Spectra Database (ASD) of the National Institute 
of Standards and Technology (NIST; \cite{NIST_ASD}).
As shown in the Table \ref{comp2}, by adding the most important $K'$ configurations 
of CV and C correlations step by step (as in Table \ref{comp1}), 
the results smoothly converge to the \textbf{CV+C RCI} results. In case when
99\% of CV and C correlations are included in the computations, the results are close to \textbf{CV+C RCI}.
The difference between the results of these calculations is only few tens of cm$^{-1}$ 
(which is a hundredth of a percent of our calculated energy levels) 
and in this case the number of CSFs 
decreases twice comparing to the space in \textbf{CV+C RCI} computations. 
By increasing the percentage 
of the CV and C correlations included in the \textbf{CV+C RCI (RSMBPT)} calculations, 
the difference between the results of \textbf{CV+C RCI (RSMBPT)} and \textbf{CV+C RCI} decreases.
The results in the 100\% column are in excellent agreement with the regular {\sc Grasp}2018 calculations. 
A difference of up to 1 cm$^{-1}$ could be due to the omission of the CV and C correlations 
with contribution $<$ than {\tt 1.0E-11} (as mentioned above, program exclude configurations with very small contribution).

Thus, it can be seen that the RCI (RSMBPT) method, when CV and C correlations are included, produces results that match the regular ones.
By selecting the most important CV and C correlations using the RSMBPT method, the CSF space can be significantly reduced.
This reduces the resources and CPU time required for RCI calculations.

Comparing \textbf{CV+C RCI} and \textbf{CV+C RCI (RSMBPT)} results in case of 100\% with NIST ASD \cite{NIST_ASD}, 
the differences for energy levels till 1,000,000 cm$^{-1}$ reach 548 cm$^{-1}$.
The disagreement of other energies is little worse, and for the level 3d$^2~^1S_0$ the difference reaches almost 2000 cm$^{-1}$.
The root-mean-square (rms) deviations obtained for all energy levels from the NIST data are 653 cm$^{-1}$.

\section{Conclusions}
The method based on the Rayleigh-Schr\"odinger perturbation theory in an irreducible tensorial form
is extended to include C correlations by providing expressions for these correlations.
This extended combination of RCI method with the RSMBPT method allows for estimation of the contribution of any $K'$ configuration of the CV and C correlation 
with preferred core and virtual orbitals for any atom or ion based on perturbation theory instead of RCI method. 
The developed method has an advantage over the regular method because it allows for selection of 
the most relevant CV and C correlations and significantly reduces the CSF space. 
This leads to a smaller matrix and makes it easier to diagonalize. 
At the same time, it reduces the resources and CPU time required for RCI calculations.
The reduction of the CSF space using the RSMBPT would be helpful and useful for calculations involving 
complex atoms and ions, as the CSF space increases rapidly when correlations 
are included in a regular way. Another important aspect in the calculations 
for complex atoms and ions is that RSMBPT method enables the CV and C correlations 
from the deeper core to be estimated and taken into account. 
The combination of the RCI and the RSMBPT methods with the advantages of RSMBPT 
extends the possibilities of the {\sc Grasp} package, particularly for the calculations of complex atoms and ions. 



\end{document}